\documentclass[aps,twocolumn,preprintnumbers,amsmath,amssymb,floatfix,groupedaddress,nofootinbib]{revtex4}

\usepackage{enumerate}
\usepackage{graphicx}
\usepackage{color}

\usepackage[applemac]{inputenc} 
\usepackage{fontenc}  
\usepackage{amsmath}

\newcommand{\beq}{\begin{equation}}
\newcommand{\eeq}{\end{equation}}
\newcommand{\bea}{\begin{eqnarray}}
\newcommand{\eea}{\end{eqnarray}}
\newcommand{\ba}{\begin{array}}
\newcommand{\ea}{\end{array}}

\newcommand{\bef}{\begin{figure}}
\newcommand{\eef}{\end{figure}}

\begin{document}

\title{Revisiting Born's rule through Uhlhorn's and Gleason's theorems.}

\author{Alexia Auff\`eves$^{(1)}$ and Philippe Grangier$^{(2)}$}

\affiliation{ 
(1): Institut N\' eel$,\;$BP 166$,\;$25 rue des Martyrs, F38042 Grenoble Cedex 9, France. \\
(2): 
Laboratoire Charles Fabry, IOGS, CNRS, Universit\'e Paris~Saclay, F91127 Palaiseau, France.}

\begin{abstract}
In  a previous article \cite{FooP}  we presented an argument to obtain (or rather infer) Born's rule, based on a simple set of axioms named ``Contexts, Systems and Modalities" (CSM).  In this approach  there is no ``emergence'', but the structure of quantum mechanics can be attributed to an interplay between the quantized number of modalities that are accessible to a quantum system, and the continuum of  contexts that are required to define these modalities.  
The strong link of this derivation with Gleason's theorem was emphasized, with the argument that CSM provides a physical justification for Gleason's hypotheses. Here we extend this result by showing that an essential one among these hypotheses - the need of unitary transforms to relate different contexts -  can be removed and is better seen as a necessary consequence of Uhlhorn's theorem.  
\end{abstract}

\maketitle

\section{Introduction.} 
\vskip -2mm

Many recent articles have proposed derivations of Born's rule  \cite{FooP,RSA,KhS,laloe,blog}, that is a major theoretical basis of quantum mechanics (QM). Let us note in particular the construction based on Quantum Darwinism, that has been proposed by Wojciech Zurek \cite{zureka,zurekb,zurekc}. It will be discussed further in the conclusion, but here we take a different position, that is  \cite{FooP}~: we start from some simple physical requirements or postulates, based on established (quantum) empirical evidence \cite{lipton,csm1,CSM_Bell,CSM_Born1,CSM_random,CSM_Born2,csm4a}; then we infer a mathematical structure able to describe these physical requirements, and finally we get deductively Born's rule and more generally the probabilistic structure of QM.  With respect to  \cite{FooP} the main purpose of the present article is to simplify further the required mathematical hypotheses, by showing that an essential one - the need of unitary transforms to relate different contexts -  can be removed and is better seen as a necessary consequence of Uhlhorn's theorem, to be introduced below. 
 \vspace{-5mm}

\section{The CSM framework.}
\vskip -2mm
The approach of  ``Contexts, Systems and Modalities" (CSM)
is a point of view on Quantum Mechanics based on a non-classical ontology, where physical properties are attributed to physical objects consisting of a system within a context, that is an idealized measurement apparatus. Such physical properties are called modalities, and a modality belongs to a specified system within a specified context, which is described classically (see Annex for more precise definitions). Loosely speaking, the mathematical description of a modality includes both a usual state vector $| \psi \rangle$, and a complete set of commuting operators admitting this vector as an eigenstate. Though it may appear heavier at first sight, this point of view eliminates a lot of troubles about QM, and can be seen (in some sense !)  as a reconciliation between Bohr and Einstein in their famous 1935 debate \cite{debatea}. 

The main feature which makes modalities non-classical is that they are both quantized and contextual, as written above. More precisely,  the empirical facts that we want to describe mathematically are :
\vskip 1mm

\noindent {\bf (i)}  in each context a measurement provides one modality among $N$ possible ones, that are mutually exclusive. No measurement can provide more than $N$ mutually exclusive modalities, and once obtained in a given context, a modality corresponds to a certain and repeatable result, as long as one remains in this same context. 
\vskip 1mm

\noindent { \bf (ii)}  the certainty and repeatability of a modality can be transferred between contexts, this fundamental property  is called  extracontextuality of modalities. All the modalities that are related together  with certainty, either in the same or in different contexts, constitute an equivalence class that we call an extravalence class. 
\vskip 1mm

\noindent {\bf (iii)}  The different contexts relevant for a given quantum system are related between themselves  by 
 transformations $g$ that have the structure of a continuous group ${\cal G}$. 
\vskip 1mm

This last statement tells that  all the different contexts relevant for a given quantum system are related between themselves 
by continuous  transformations $g$ which  are associative, have a neutral element (no change), and an inverse.
Therefore this set  has the structure of a continuous group ${\cal G}$, which 
is generally not commutative (such as the rotations of a macroscopic device).  
Our goal is then to identify a (non-classical) probabilistic framework \cite{FooP} corresponding to these requirements, and to 
draw consequences by using suitable standard theorems. 
\vskip 1mm

For this purpose, the central mathematical ingredient is to associate a rank-one projector $P_i$ (a $N \times N$ hermitian matrix such as $P^2 = P = P^\dagger$ ) to each modality, with the rule that modalities associated with orthogonal projectors are mutually exclusive, and modalities associated with the same projector are mutually certain. Correspondingly, a context is associated with a set of rank-one mutually orthogonal projectors, whereas an extravalence class of modalities is associated with a single projector. 
 In addition, we assume  that, given a modality in a context, the probability to get another modality in another context is a function of the two projectors associated with these two modalities (or equivalently with their two extravalence class).

The heuristic motivation for using a complete set of mutually orthogonal projectors to build up a context is that this ensures that the events associated with modalities cannot be subdivided in more elementary events, as this would be the case with classical (partition-based) probabilities. On the other hand, the construction warrants that  certainty can be transferred between contexts for extravalent modalities. 

Now, we want to show that the usual structure of QM follows from the above hypotheses : this means that unitary transforms between projectors as well as Born's rule are necessary in the above framework.  Let us emphasize that it is easy to show these results fulfill our hypotheses; but showing that they are necessary requires powerful (and difficult to demonstrate) mathematical theorems. Necessity also means that if one wants to give up unitary transforms or Born's rule, one has to give up one of the statements above, without contradicting empirical evidence; this is an interesting challenge \cite{challenge}. 
\vspace{-3mm}

\section{Necessity of unitary transforms.} 
\vskip -2mm

As said above, the basic mathematical tool we use is to associate  $N$ mutually orthogonal projectors with  the $N$ mutually exclusive modalities within a given context. 
The choice of such a specific orthogonal set of projectors associated with a context is not given a priori, but once it is done, the sets of projectors in all other contexts should be obtained by a bijective map $\Gamma$ reflecting the structure of the continuous group ${\cal G}$ of context changes. 

For consistency, 
if two orthogonal projectors are associated with two mutually exclusive modalities, they should stay orthogonal under the map  $\Gamma$,
whatever choice is made for the projectors associated with a ``reference" (fiduciary) context.  Then let us consider 
\vskip 2mm

{\bf Uhlhorn's theorem \cite{uhlhorn,chevalier} :} Let ${\cal H}$ be a complex Hilbert space with  $dim({\cal H})  \geq 3$, and let $P_1({\cal H})$ denote the set of all rank-one projections on ${\cal H}$.  Then every bijective map $\Gamma : P_1 ({\cal H} ) \rightarrow P_1 ({\cal H} )$, such that  $pq = 0$ in $P_1({\cal H})$  if and only if $\Gamma(p) \Gamma(q) = 0$, is induced by a unitary or anti-unitary operator on the underlying Hilbert space.
\vskip 1mm

This theorem implies that if orthogonality is conserved as required above, then 
the transformations between the sets of projectors associated  with different contexts is unitary or anti-unitary \cite{au}.  In the case of a continuous group of transformations, which is the case here, then the transformation must be unitary (and not anti-unitary) as long as  it is continuously connected to the identity, which is the situation we are interested in (see also below). 
\vskip 1mm

The strength and importance of Uhlhorn's theorem is that it requires that the map keeps the orthogonality of rank-one projections, or equivalently of non-normalized vectors (or rays). A  transformation mapping an orthonormal basis onto an orthonormal basis 
is clearly a unitary transform;  but this result is far from obvious if the conservation of the norm is not required. 
A related (but weaker) result is Wigner's theorem, getting the same conclusion as Uhlhorn's  if the modulus of the scalar product of any two vectors is conserved by the transformation. Uhlhorn's theorem is much more powerful, since it only assumes that  the scalar product is conserved  when it is zero, i.e. when the two rays are orthogonal \cite{Semrl}. 
\vskip 1mm

We thus get a major result : once a set of mutually orthogonal projectors associated with a fiduciary context has been chosen, the sets of projectors associated to all other contexts are obtained by unitary transformations, so we are unitarily ``moving" in a Hilbert space.  
There are also various arguments for using unitary (complex) rather than orthogonal (real) matrices; in our framework the simplest argument is to require that all permutations of modalities within a context are continuously connected to the identity. This is not possible with (real) orthogonal  matrices, which split into two subsets with determinants $\pm 1$,  but is possible with unitary ones, 
see \cite{CSM_Born1,CSM_Born2}. 
\vspace{-4mm}

\section{Necessity of Born's rule}
\vskip -2mm

The next step is to consider the probability $f(P_i)$ to get a modality associated with projector $P_i$.  By construction a context is such that $\sum_{i=1}^{i=N}  P_i = I$, and $\sum_{i=1}^{i=N}  f(P_i) = 1$ for any complete set $\{ P_i \}$.  But these are just the hypothesis of Gleason's theorem, 
so there  is a density matrix $\rho$ such that $f(P_i) = \textrm{Trace}(\rho P_i)$.  More precisely :
\vskip 2mm

{\bf Gleason's Theorem  \cite{gleason,cooke}~: } Let $f$ be a function to the real unit interval from the projection operators on a separable (real or complex)  Hilbert space  with dimension at least 3. If one has $\sum_i  f(P _{i}) = 1$ for any set  $\{P_{i} \}$ of mutually orthogonal rank-one projectors summing to the identity, then there exists a positive-semidefinite self-adjoint  operator $\rho$ with unit trace (called a density operator) such that $f(P_i) = \textrm{Trace}(\rho P_i)$. 
\vskip 2mm

If the value 1 is reached, then  $\rho$ is also a projector $Q_j$ and $f(P_i) = \textrm{Trace}(Q_j  P_i )$ which is the usual Born's formula. 
As already explained in  \cite{FooP}, we have considered initial and final modalities, i.e. rank 1 projectors \cite{CSM_Born2}, but more generally Gleason's theorem  provides the probability law for density operators (convex sums of projectors), interpreted as statistical mixtures. This clarifies the  link between Born's rule and the mathematical structure of density operators \cite{Masanes}. One gets thus the basic probabilistic framework  of QM; this is enough for our purpose here, but more is needed for a full reconstruction;
in particular, composite systems and tensor products should be included \cite{completinga}. 
\vskip 2mm

In addition, one must define explicitly the relevant physical properties and associated contexts, that may go from space-time symmetries (Galileo group, Lorentz group) to qubits registers. Then the unitary transforms appear as representations of the relevant group of symmetry  \cite{book}. In any case, contextual quantization applies and sets the scene where the actual physics takes place. 

\vspace{-4 mm}
\section{Discussion.} 
\vskip -2mm

For the sake of completeness, it is useful to remind here some statements already presented in \cite{FooP}. A key feature of the contextual quantification postulate is the fixed value N of the maximum number of mutually exclusive modalities, that  turns out to be the dimension of Hilbert space. This provides another heuristic reason for using projectors: the projective structure of the probability law guarantees that  the maximum number of mutually exclusive modalities cannot be circumvented. 

This would not be the case in the usual partition-based probability theory: partitioning all modalities into N subsets for any given context would not prevent subpartitions, corresponding to additional details or hidden variables forbidden by our basic postulate. This is mathematically equivalent to the Bell's  or Kochen-Specker's (KS) theorems and all their variants, which essentially demonstrate the inadequacy of probabilities based on partitions.  This problem disappears when projectors are used, and then, starting from Gleason's theorem, there is no choice but Born's rule. 

It should also be noted that Bell's or KS theorems consider discrete sets of contexts, while Gleason's theorem is based on the interaction between the continuum of contexts and the quantified number of accessible modalities, in a given context. This feature is also fully consistent with the ideas of CSM.  Therefore, Gleason's assumptions in our approach have a deep physical content that combines contextual quantification and extracontextuality of modalities. Since these features are required by empirical evidence, the usual QM formalism provides a good answer to a well-posed question.

We note however that our approach leads to some differences with the standard (textbook) one: in particular, the usual quantum state vector  $|\psi \rangle$ is not predictively complete, since it provides a well-defined probability distribution only when ``completed" by the specification of a context \cite{inference}. A complete description including also the contexts requires the use of algebraic methods \cite{completinga}. 

To conclude, let us come back to some epistemological difference between the approach used here and the one favored  by Wojciech Zurek \cite{zureka,zurekb,zurekc}. In his point of view, the role of mathematics is prescriptive : first ``Let be $\Psi$'', and then all the rest should follow. On the contrary,  in our approach its role is descriptive : there is a physical world out there, and the mathematical langage is our  best tool  to  ``speak" about it --  but  it is a langage, not the Tables of the Law. Also, in CSM there is no ``Emergence of the Classical" \cite{zurekc}: the classical and quantum description are both needed to make sense of our physical universe, where an object is a system within a context.  These subtle differences may appear more philosophical than practical, and they do not preclude an agreement on more down-to-the-earth issues, e.g.  the management of decoherence for applications to quantum technologies. However, keeping such issues open is certainly a compost for new ideas to germinate. 
\vskip 3mm

{\bf Acknowledgements:}
The authors thank  Franck Lalo\"e and Roger Balian for many useful discussions, 
and Nayla Farouki for continuous support.

\vskip 5mm
{\centerline{\large \bf Annex : CSM definitions and postulates.}}
\vskip 2mm

For the convenience of the reader, this Appendix summarizes the basic elements of CSM, 
see e.g. \cite{FooP}.
\vskip 2mm

{\bf Postulate 1a (ontology)  
 : } {\it Let us consider a quantum system S interacting with a specified set of measurement devices, that is called a {\bf context}. 
The best physically allowed measurement process provides a set of numbers, corresponding to the values of a well-defined and complete set of jointly measurable quantities. Ideally these values will be found again with certainty, as long as the system and context are kept the same; they define a {\bf modality}, belonging to a system within a context \cite{tt}. }
\vskip 1mm

Here the word ``context" includes the actual  settings of the device, e.g. the fact that $S_z$ is measured rather than $S_x$: the context must be factual, not contrafactual. On the other hand all devices designed to measure $S_z$ are equivalent as a context, in a (Bohrian) sense that they all define the same conditions for predicting the future behaviour of the system. Note that the modalities are not defined in the same way as the usual ``quantum states of the system", since they are explicitly attached to both the context and the system. This leads to the addition : 
\vskip 2mm

\noindent  {\bf Postulate 1b (extravalence) : }
{\it When S interacts in succession with different contexts, certainty and repeatability may be transferred between their modalities. This is called {\bf extracontextuality}, and  defines an equivalence class between modalities, called {\bf extravalence}.}
\vskip 1mm

\noindent The equivalence relation is obvious, for more details and examples of extravalence classes  see  \cite{CSM_Born2}. Note that extravalent modalities appear only if $N \geq 3$, this has an obvious geometrical interpretation in relation with both Gleason's and Uhlhorn's theorems.

From the above postulates, 
 one measurement provides one and only one modality. Therefore in any given context the various possible modalities are mutually exclusive, meaning that if one is true, or verified, all other ones are not true, or not verified. This is formalized by  
\vskip 2mm

\noindent {\bf Postulate 2 (contextual quantization)} : {\it 
For a given context, i.e. a given ``knob settings" of the measurement apparatus,
there exist $N$ modalities  that are mutually exclusive. 
The value of $N$, called the dimension, is a characteristic 
property of a given quantum system, and is the same in any relevant context.}  
 
Modalities observed in different contexts are generally not mutually exclusive, they are said to be  incompatible,
meaning that if a result is true, or verified, one cannot tell whether the other one is true or not. 
Finally, a last statement defines the relation between contexts :
\vskip 2mm

\noindent {\bf Postulate 3  (changing contexts) :}  {\it The different contexts relevant for a given quantum system are related between themselves  by (classical) transformations $g$ that have the structure of a continuous group ${\cal G}$. }
\vskip 2mm

The intuitive idea behind these statements  is that making more measurements in QM (by changing the context) cannot  provide ``more details" about the system, because this would increase the number of mutually exclusive modalities, contradicting Postulate 2. One might conclude that changing context randomizes all results, but this is not true either~: some modalities may be related with certainty between different contexts. 
%
This is why extravalence is an essential feature of the construction, both as a physical requirement, and as a justification for Gleason's hypotheses. Adding that context changes must preserve the mutual exclusiveness of modalities, i.e. must preserve the orthogonality of projectors, that is Uhlhorn's hypotheses, makes Born's rule a necessity. 




\begin{thebibliography}{999}
\bibitem{challenge} In fact ``universal" unitary equivalence does not hold at the context level,  because the contexts are classical (there is no quantum superposition of contexts).  This requires to complete QM using algebraic methods  \cite{completinga}.
\bibitem{au}  An anti-unitary operator  $U$ is a bijective antilinear map such that $\langle Ux | Uy \rangle = \langle x | y \rangle^*$ for all vectors  $x$,  $y$ in $\cal H$.
\bibitem{tt} The free evolution of the system is omitted here; if it is present, the result of a new measurement can still be predicted with certainty, but in another context that can be deduced from the free evolution.
\end{thebibliography}
\end{document}